\documentclass[intlimits,twoside,a4paper]{article}

\usepackage[cp1251]{inputenc}
\usepackage[eqsecnum]{cmpj3}
\usepackage{bm}
\usepackage{multirow}

\issue{2021}{24}{2}{23706}
\doinumber{10.5488/CMP.24.23706}
\title[Band structure of 2H-SiC and 4H-SiC]%
{Theoretical study of the band structure of 2H-SiC and 4H-SiC of silicon carbide polytypes}%

\author[A.~V. Sinelnik, A.~V. Semenov]{A.~V. Sinelnik\orcid{0000-0002-5045-687X}, A.~V. Semenov\orcid{0000-0001-8663-065X}}
\address{National Technical University ``Kharkiv Polytechnic Institute'', 2, Kyrpychova str., Kharkiv, 61002, Ukraine}

\date{Received July 16, 2020, in final form March 18, 2021}

\begin{document}

\maketitle

\begin{abstract}
We have studied the electronic band properties of 2H-SiC and 4H-SiC silicon carbide polytypes. 
The structures of the electronic bands and density of state (DOS) using ab initio Density 
Functional Theory (DFT) were calculated for the first Brillouin zone both in the generalized 
gradient approximation and taking into account quasiparticle effects according to the GW scheme. 
The calculated bandgaps obtained using the GW approximation ${E_{\rm{g~2H-SiC}} = 3.17}$~eV and 
${E_{\rm{g~4H-SiC}} = 3.26}$~eV agree well with experimental values. The shape and values of total DOS 
are within agreement with calculations performed by other authors. The calculated total energy 
values for 2H-SiC and 4H-SiC were close, but they satisfy the condition ${E_{\rm{2H}} > E_{\rm{4H}}}$, which 
implies that the 4H-SiC structure is more stable than 2H-SiC. Our calculations of the band 
structure and DOS of 2H-SiC and 4H-SiC by the DFT method showed that the application of the 
GW approximation is an optimum approach to the study of the electronic structure of 2H-SiC and 
4H-SiC polytypes.
%
\end{abstract}

\section{Introduction}

Silicon carbide (SiC) is one of the most promising materials for high-temperature, 
radiation-resistant, power and high-speed electronics, as it has unique physical 
and electronic properties. These properties include: a wide band gap (approximately 
three times larger than that of silicon), a high critical avalanche breakdown field 
(approximately 10 times greater than that of silicon), and a high electron drift 
saturation rate (2.5 times greater than in silicon and gallium arsenide), high 
thermal stability, chemical inertness, excellent mechanical properties, etc.~\cite{MorkocJAP1,Saddow2,Harris3,Roccaforte4}. 
Silicon carbide exists in many polytype modifications (more than 250 have been established~\cite{Cheung5}), 
which are derivatives of hexagonal and rhombohedral close-packed crystal lattices~\cite{Verma6}. 
All silicon carbide polytypes crystallize according to the laws of close ball packing 
and are binary structures built of identical layers, differing both in the order of 
arrangement of the cubic or hexagonal layer, and in the number of these layers in the 
unit cell. Despite the fact that studies of the structural and electronic properties 
of SiC polytypes have been carried out  for more than 100 years, there 
is still no unified approach to describing the dependence of polytype properties on 
their degree of hexagonality~\cite{Verma6}, and there is a scatter of theoretical and 
experimental data on the main values of the electronic parameters of polytypes~\cite{Pizzagalli7,Xuejie8,Capitani9,Azadegan10,Persson11,Zhao12,Son13,Kaczer14}. 
The most widespread and, accordingly, the most studied are 4H-SiC, 6H-SiC, 3C-SiC, 15R-SiC polytypes. 
This is due to the technological capabilities of obtaining single crystals of 
these polytypes. Obtaining other SiC polytypes, for example, such as 2H-SiC, is difficult, 
due to the difficulty of providing stable conditions that prevent the growth of 
energetically more favorable polytypes (for example, 3C-SiC, 4H-SiC, 6H-SiC). In this 
regard, in the literature there are very little data from experimental studies of this SiC 
polytype. In addition, there are quite large differences in the results of theoretical studies. 
At the same time, the properties of 2H-SiC are of great interest, due to its largest 
band gap among SiC polytypes~\cite{Harris3}. It is advisable to carry out the theoretical analysis 
of the electronic properties of 2H-SiC in comparison with other polytypes with established 
experimental data. The most convenient is a tandem with the well experimentally measured 
4H-SiC polytype, the unit cell of which is two times longer than in 2H-SiC~\cite{Cheung5}.

In this work, we presented the results of first-principles calculations based on the DFT 
band electronic properties of 2H-SiC and 4H-SiC in comparison with the results of the studies 
obtained by other authors.

\section{Computational methods}

The calculation of the energy band structure of electrons from the first principles 
in this work was carried out in the framework of the density functional theory 
(DFT) using the software package \emph{exciting}~\cite{Gulans15}. As the electron wave functions, 
this package utilizes the approximation of linearized augmented plane waves (LAPW) 
and localized orbitals (LO) as basis functions. The exchange-correlation component 
of the potential was chosen in the form of  asymptotically corrected 
generalized gradient approximation (acGGA) using the Burke functional~\cite{Burke16}.
The acGGA functional is chosen for its accurateness in minimizing atomic and energy 
errors for large-$Z$ atoms and it is pure first-principle (not relying on empirical data). 
The first circumstance is key in the procedure used for optimization of atomic positions.
Since within the framework of LDA 
(local density approximation) and GGA-DFT, it is not possible to obtain accurate 
values of the band gap for semiconductor and dielectric materials~\cite{Yakovkin17}. We 
calculated the band structure taking account of quasiparticle (QP) effects using 
the GW approximation of the electron self-energy that takes quasiparticle 
effects into account~\cite{Hedin18}. We used the single-shot GW (also known as $\rm{G}_0\rm{W}_0$) 
approximation. To obtain the starting position for GW calculations, the band 
structure was first calculated according to the GGA-DFT scheme. The self-consistent 
calculations GGA-DFT were performed until the convergence condition was reached: 
the absolute change in the total energy and effective potential is less than 
$10^{-6}$~eV per cell. The lattice was relaxed using the method of 
Broyden-Fletcher-Goldfarb-Shanno~\cite{Byrd19}, while the unit cell dimensions remained 
constant, and only atomic positions varied.

\section{Results and discussion}

\begin{figure}[htbp]
\vspace{-2ex}
\centerline{\includegraphics[width=0.80\columnwidth]{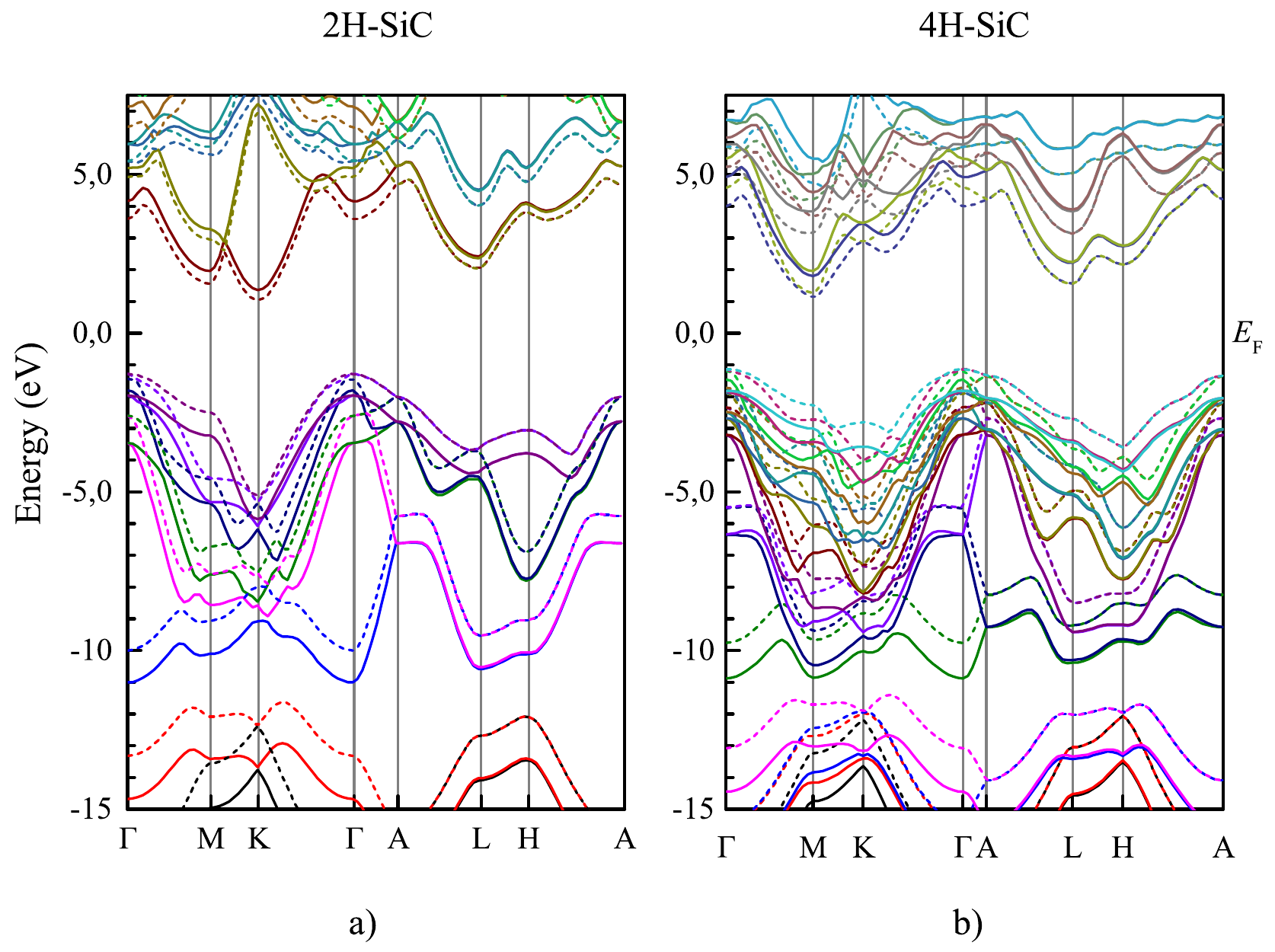}}
\caption{(Colour online) Band structure diagrams for 2H-SiC (a) and 4H-SiC (b) 
obtained using GGA-DFT (dashed lines) and taking into account 
the quasiparticle effects of GW-DFT (solid lines).}
\label{Fig1}
\end{figure}

The results of the band structure calculations of 2H-SiC and 4H-SiC crystal 
structures using the GGA-DFT functional and subsequently using the correction 
based on GW approximation of self-energy are presented in figure~\ref{Fig1}. 
The corresponding energies of the band structure at the points of high 
symmetry near the band gap are given in table~\ref{Tbl1}. 

The electron energy bands for 2H-SiC and 4H-SiC in the energy range from $-15$ to 
$+8$~eV along the directions of high symmetry in the first Brillouin zone are shown 
in figure~\ref{Fig1}~(a) and~(b), respectively. The energy in the diagrams is 
relative to the Fermi energy level. The occupied states of the 4H-SiC valence band 
calculated by us, figure~\ref{Fig1}~(b) are close to the previously published results 
of ab initio LAPW-LDA~\cite{Persson11}, LCAO (linear combination of atomic orbitals)-LDA~\cite{Zhao12} 
and PW (plane wave)-GGA~\cite{Ming24} calculations. The minimum of the conduction band is 
at point K of Brillouin zone for 2H-SiC, which is similar to the theoretical result~\cite{Pizzagalli7,Ming24}. For 4H-SiC in our calculations, the minimum of the conduction band is at the
point M. At point M the second minimum of the conduction band is observed, which is 
only 0.13~eV and 0.16~eV higher than the lowest unoccupied state in the GGA and GW 
approximations, respectively. These results are consistent with the ballistic electron 
emission microscopy study~\cite{Kaczer14}, in which the second minimum was observed at about 
0.15~eV, as well as with other ab initio calculations, in which it is at 0.12~eV~\cite{Persson11} 
and 0.18~eV~\cite{Zhao12}. In our quasiparticle band structure calculations, the split-off 
band and the bands of light and heavy holes overlap at the point $\Gamma$. There 
is no overlap on the GGA band structures. This overlap maximizes at the point 
$\Gamma$ and reaches 0.17~eV and 0.36~eV for 2H-SiC and 4H-SiC, respectively. The 
calculated indirect band gap with QP correction using GW scheme according to our 
data is 3.17~eV for 2H-SiC (see table~\ref{Tbl1}), which is in good agreement with the 
experimental data of 3.3~eV~\cite{Harris3}. This value differs from the theoretical results 
of the 2H-SiC calculations performed by other authors: 3.68~eV (QP-GW calculation)~\cite{Wenzien20}, 3.75~eV (self-consistent QP-GW)~\cite{gao23}, 3.01~eV~\cite{Ming24} and is very close in other QP-GW calculations 3.15~eV~\cite{ummels22}, 3.12~eV~\cite{gao23}. Indirect band gap in our calculated electron band structure of 
4H-SiC is 3.26~eV, which is comparable with the theoretical data of 3.56~eV (QP-GW calculations)~\cite{Wenzien20}, close to 3.35~eV (QP-GW calculations)~\cite{ummels22} and differs from the DFT-GGA calculations in~\cite{Ming24}, 
which reported 2.45~eV. The experimental value of the band gap in 4H-SiC is 3.29~eV~\cite{Harris3}. 
A significant difference between our values of the band gap from the values obtained 
in the LDA and GGA approximations is due to the known limitations of this approach. 
Obviously, the use of quasiparticle correction of the self-energy of the electron allows 
us to obtain the values of the band gap that are in good agreement with the values 
measured experimentally.
\begin{figure}[htbp]
\vspace{-2ex}
\centerline{\includegraphics[width=0.85\columnwidth]{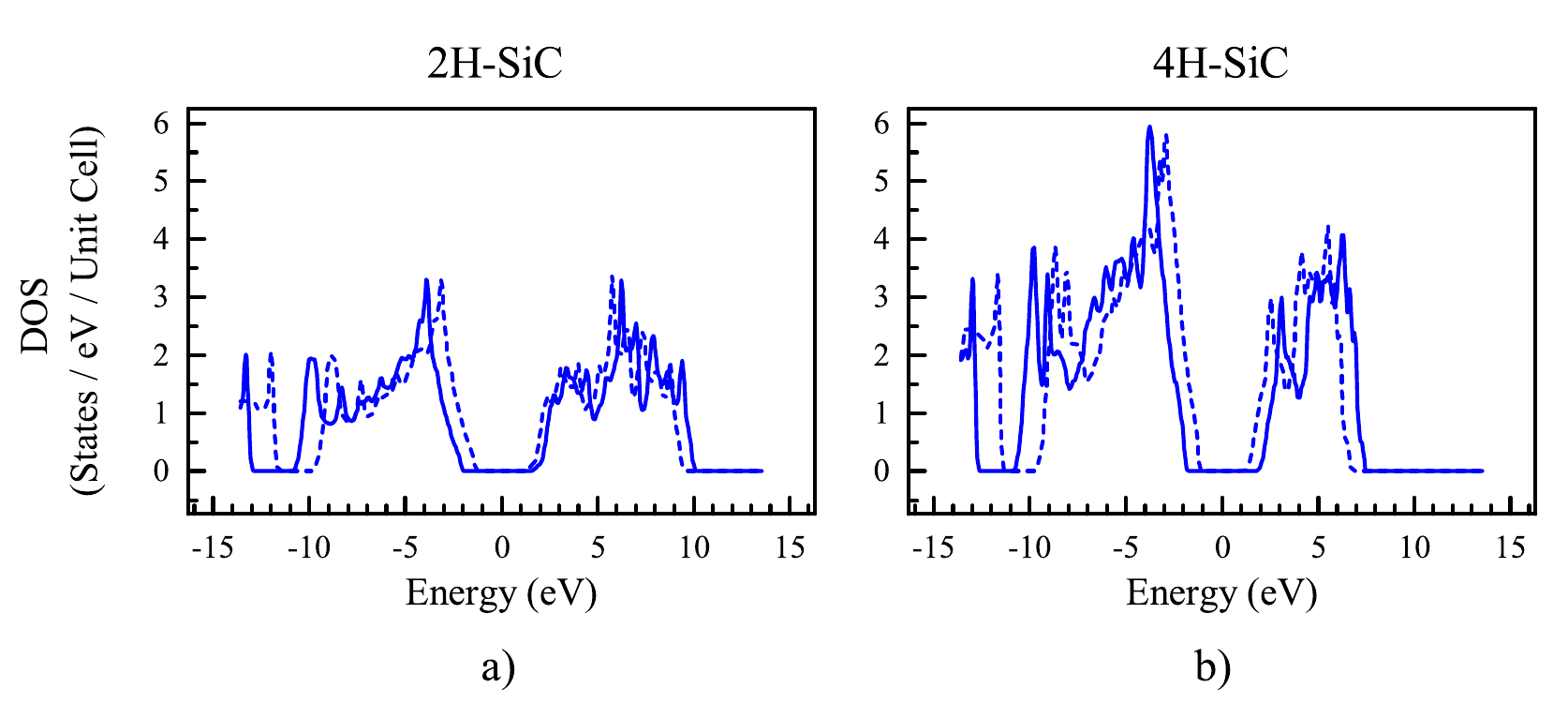}}
\caption{(Colour online) The total density of states for 2H-SiC (a) and 4H-SiC (b) 
obtained using GGA-DFT (dashed lines) and taking into account the 
quasiparticle effects using GW-DFT (solid lines).}
\label{Fig2}
\end{figure}

The calculated total density of states of 2H-SiC and 4H-SiC is shown in 
figure~\ref{Fig2}~(a) and~(b), respectively. The shape of the DOS located below $-10$~eV 
has similar form for both of the considered polytypes, while those states that 
are near the edge of the conduction band are significantly different (figure~\ref{Fig2}). 
The slope of the DOS at the edge of the conduction band becomes steeper for the 
4H-SiC structure. The difference between the DOS form according to the results 
of GGA and GW calculations is insignificant: only a relative shift of the bands 
along the energy axis is observed. The value of DOS in the entire energy range for 
4H-SiC is on average higher than that in 2H-SiC, which agrees with the recent 
DFT-GGA calculations~\cite{Ming24}.

\begin{table}[!t]
\caption{The energy values of the band structure (eV) at the points of high symmetry of 
the Brillouin zone, as well as the band gap values in the GGA-DFT approximation and 
in the GW approximation. The band gap values obtained from the results of theoretical 
and experimental studies of other authors are also presented.}
\label{Tbl1}
\vspace{2ex}
\begin{center}
\begin{tabular}{|l|c|c|c|c|}
\hline
\multicolumn{1}{|c|}{\multirow{2}{*}{Band state}}                                                       & \multicolumn{2}{|c|}{2H-SiC}                                                       & \multicolumn{2}{|c|}{4H-SiC}                                                       \\ \cline{2-5} 
\multicolumn{1}{|c|}{}                                                                          &              GGA                         &                 GW                   &                    GGA                 &                  GW                     \\ \hline
$\Gamma_{\rm{6v}}$                                                                                           & 0                                     & $-0.17$                                 & 0                                     & $-0.36$                                 \\ \hline
$\Gamma_{\rm{1v}}$                                                                                          & $-0.15$                                 & 0                                     & $-0.06$                                 & 0                                     \\ \hline
$\Gamma_{\rm{1c}}$                                                                                          & 4.89                                  & 5.96                                  & 5.36                                  & 6.59                                  \\ \hline
$\rm{M_{4v}}$                                                                                           & $-1.19$                                 & $-1.42$                                 & $-1.10$                                 & $-1.54$                                 \\ \hline
$\rm{M_{1c}}$                                                                                           & 2.87                                  & 3.76                                  & 2.30                                  & 3.26                                  \\ \hline
$\rm{K_{2v}}$                                                                                           & $-3.81$                                 & $-4.04$                                 & $-1.66$                                 & $-2.11$                                 \\ \hline
$\rm{K_{2c}}$                                                                                           & 2.33                                  & 3.17                                  & 4.03                                  & 4.91                                  \\ \hline
$\rm{H_{3v}}$                                                                                           & $-1.77$                                 & $-1.98$                                 & $-2.44$                                 & $-2.84$                                 \\ \hline
$\rm{H_{3c}}$                                                                                           & 5.08                                  & 5.86                                  & 3.32                                  & 4.19                                  \\ \hline
$\rm{A_{5,6v}}$                                                                                         & $-0.70$                                 & $-0.96$                                 & $-0.19$                                 & $-0.59$                                 \\ \hline
$\rm{A_{1,3c}}$                                                                                         & 5.98                                  & 7.08                                  & 5.36                                  & 6.59                                  \\ \hline
$\rm{L_{1,2,3,4v}}$                                                                                     & $-2.39$                                 & $-2.68$                                 & $-1.54$                                 & $-1.92$                                 \\ \hline
$E_{\rm{g}}$ (direct)                                                                                   & 4.06 (M)                              & 5.18 (M)                              & 3.40 (M)                              & 4.80 (M)                              \\ \hline
$E_{\rm{g}}$ (indirect)                                                                                 & 2.33 ($\Gamma$-K)                            & 3.17 ($\Gamma$-K)                            & 2.30 ($\Gamma$-M)                            & 3.26 ($\Gamma$-M)                            \\ \hline
\multicolumn{5}{|l|}{} \\ \hline
$E_{\rm{g}}$ (GW \cite{Wenzien20})                                                                                 & \multicolumn{2}{|c|}{3.68}                                                       & \multicolumn{2}{c|}{3.56}                            \\ \hline
$E_{\rm{g}}$ (PBE-GGA \cite{Nuruzzaman21})                                                                                 & \multicolumn{2}{|c|}{3.01}                                                       & \multicolumn{2}{c|}{2.45}                            \\ \hline

$E_{\rm{g}}$ (GW \cite{ummels22})                                                                                & \multicolumn{2}{|c|}{3.15}                                                       & \multicolumn{2}{c|}{3.35}                            \\ \hline

$E_{\rm{g}}$ (GW \cite{gao23})                                                                                 & \multicolumn{2}{|c|}{3.12}                                                       & \multicolumn{2}{c|}{}                            \\ \hline

$E_{\rm{g}}$ (QPscGW \cite{gao23})                                                                                 & \multicolumn{2}{|c|}{3.75}                                                       & \multicolumn{2}{c|}{}                            \\ \hline

\multicolumn{5}{|l|}{} \\ \hline
$E_{\rm{g}}$ (exp. \cite{Harris3})                                                                             & \multicolumn{2}{c|}{3.3}                                                       & \multicolumn{2}{c|}{3.29}   \\ \hline                                                
\end{tabular}
\end{center}
\end{table}

\section{Conclusion}

The total energy of the 2H-SiC and 4H-SiC polytypes per atom was 
calculated for relaxed crystal lattices optimized during the 
self-consistent calculations (table~\ref{Tbl2}). The calculated 
values of the total energy of both structures are very close to 
each other. However, the total energy ratio satisfies the condition 
$E_{\rm{2H}} > E_{\rm{4H}}$, which means that the 4H-SiC structure 
is more stable than 2H-SiC, which is consistent with the known 
theoretical and experimental data~\cite{MorkocJAP1,Saddow2,Harris3}. 
For comparison, the table shows the total energy for the 2H-SiC and 
4H-SiC structures from~\cite{Nuruzzaman21,Ming24}.

\begin{table}[htbp]
\caption{The calculated total energy values per atom (eV) of 2H-SiC and 
4H-SiC structures as well as results from other studies.}

\label{Tbl2}
\vspace{2ex}
\begin{center}
\begin{tabular}{|l|l|l|}
\hline
               & 2H-SiC & 4H-SiC \\ \hline
$E_{\rm{tot}}$ (present) & $-7.388$ & $-7.392$ \\ \hline
$E_{\rm{tot}}$ \cite{Nuruzzaman21}        & $-9.659$ & $-9.660$ \\ \hline
$E_{\rm{tot}}$ \cite{Ming24}        & $-7.396$ &        \\ \hline
\end{tabular}
\end{center}
\end{table}

We studied the electronic properties of 2H-SiC and 4H-SiC of silicon carbide 
polytypes. The structures of the electronic bands and DOS 2H-SiC and 4H-SiC 
using ab initio DFT were calculated for the first Brillouin zone both in the 
GGA-DFT approximation and taking into account quasiparticle effects according 
to the GW scheme. The calculated bandgaps obtained using the GW approximation, 
${E_{\rm{g~2H-SiC}} = 3.17}$~eV, ${E_{\rm{g~4H-SiC}} = 3.26}$~eV 
agree well with the known theoretical and experimental values. The shape and 
values of total DOS are within agreement with calculations performed by other 
authors. The calculated total energy values for 2H-SiC and 4H-SiC are close, 
but they satisfy the condition ${E_{\rm{2H}} > E_{\rm{4H}}}$, which implies that 
the 4H-SiC structure is more stable than 2H-SiC. Our calculations of the band 
structure and DOS of 2H-SiC and 4H-SiC by the DFT method showed that the 
application of the GW approximation is an optimum approach for studying the 
electronic structure of 2H-SiC and 4H-SiC polytypes.

%
%
%
%

\newpage
\ukrainianpart

\title{Теоретичне дослідження зонної структури 2H-SiC та 4H-SiC політипів карбіду кремнія}
\author{О.~В. Синельник, О.~В. Семенов}
\address{Національний технічний університет “Харківський політехнічний інститут”, вул. Кирпичова, 2, Харків, 61002, Україна}

\makeukrtitle

\begin{abstract}
\tolerance=3000%
Ми провели дослідження властивостей електроних зон 2H-SiC and 4H-SiC політипів карбіду кремнія. 
Були проведені ab initio розрахунки зонної структури та щільності станів для першої зони Брилюена  з 
використанням теорії функціоналу щільності, як у рамках апроксимації узагальненого градієнту, так із 
врахуванням квазічастинкових ефектів по схемі GW. 
Розраховані ширини забороненої зони з використанням GW апроксимації ${E_{\rm{g 2H-SiC}} = 3.17}$~еВ та 
${E_{\rm{g 4H-SiC}} = 3.26}$~еВ добре узгоджуються з експериментальними значеннями. Форма та значення повної щільності станів узгоджуються з обчисленнями інших авторів. Розраховані значення повної енергії 
2H-SiC та 4H-SiC є близькими, але вони задовільняють умові $E_{\rm{2H}} > E_{\rm{4H}}$, що 
означає, що структура 4H-SiC більш стабільна ніж 2H-SiC. Наші обчислення зонної структури та щільності станів 2H-SiC та 4H-SiC методом теорії функціоналу щільності показало що застосування апроксимації 
GW є оптимальним для вивчення електронної структури політипів 2H-SiC та 
4H-SiC.

\end{abstract}

\end{document}